# Describe NMR relaxation by anomalous rotational or translational diffusion

Guoxing Lin*

*Carlson School of Chemistry and Biochemistry, Clark University, Worcester, MA 01610, USA*

*Email: glin@clarku.edu

**Abstract**

This paper employs the general time-space fractional diffusion equation $\{0 < \alpha, \beta \leq 2\}$ to derive correlation time function for analyzing nuclear magnetic resonance (NMR) relaxation. Both the anomalous rotational and translational diffusion are treated. NMR relaxation time affected by various Hamilton interactions such as dipolar or quadrupolar couplings can be calculated from the Mittag-Leffler type time correlation and their corresponded spectral density functions obtained. Additionally, to verify the results, the theoretical expressions are applied to fit reported experimental data of NMR quadrupolar coupling relaxation of head-to-head poly(propylene) (hhPP) in a polymer blend. The fitting is excellent and more convenient than the fitting utilising the traditional modified Kohlrausch-Williams-Watts (KWW) formalism. Further, it is found that the temperature dependence behavior of the segmental dynamics in anomalous diffusion may obey a different Vogel-Tamman-Fulcher (VTF) expression. The paper proposes new, general formalisms for analyzing various NMR relaxation experiments in macromolecular systems.

## 1. Introduction

Nuclear magnetic resonance (NMR) has been a powerful tool for monitoring molecular motions [1,2] in many systems such as macromolecular systems [3,4,5]. Molecular motions alter relative molecular orientations, which modulates many fundamental Hamiltonian interactions of NMR spin systems including dipolar coupling, quadrupolar coupling, chemical shift anisotropy, etc.[1] The relaxation process of NMR spin systems are affected by the random fields arising from these modulated interactions causing the spin transitions to take place. When the molecular motion frequency is resonant with the transition frequency of the spin system, effective transition creates a fast NMR relaxation. In contrast, an off-resonance molecular motion leads to a slow NMR relaxation. The rate of NMR relaxation can usually be calculated based on the time correlation function of the rotational diffusion or translational diffusion.

The type of time correlation function for NMR relaxation depends on the motion of the investigated spin systems. A small molecule undergoing a normal rotational or translational diffusion in a liquid solution often has a mono-exponential time correlation function [1]. However, a mono-exponential time correlation function is often insufficient in describing the molecular motion in many complex systems such as macromolecular systems. The Kohlrausch-Williams-Watts (KWW) function [6,7] is the commonly used non-mono-exponential correlation distribution function for macromolecular systems. The function is a stretched exponential function $\exp\left(-\left(\frac{t}{\tau}\right)^\alpha\right)$ where $\alpha$ is the stretching exponential constant, and $\tau$ is the characteristic relaxation time. Mittag-Leffler function $E_{\alpha,1}\left(-\left(\frac{t}{\tau}\right)^{\alpha}\right)$ is another type of function that has been employed to describe the correlation function for NMR relaxation and other fractional relaxation processes [8, 9,10,11,12]. Additionally, the Mittag-Leffler function $E_{\alpha,1}(-x)$ is approximately the same as the stretched exponential function at small $x$ values, but at large $x$ values, it decreases more slowly than the stretched exponential function.

Both the stretched exponential type correlation function and Mittag-Leffler type correlation function can be explained by the anomalous diffusion [13, 14,15, 16,17], which has time derivative order $\alpha$ and space derivative order $\beta$. Based on the values of $\alpha$ and $\beta$, the fractional diffusion can be classified into the following three categories: the time-fractional diffusion with $\{0 < \alpha, \leq 2, \beta = 2\}$, the space-fractional diffusion with $\{\alpha = 1, 0 < \beta \leq 2\}$, and the general fractional diffusion with $\{0 < \alpha, \beta \leq 2\}$. From the perspective of the continuous-time random walk (CTRW), the anomalous diffusions arise because of the jump waiting time distribution that behaves asymptotically to an inverse power law, or the jump length distribution following a power law [18,19]. These anomalous diffusions have been applied to dynamics study by relaxation NMR or pulsed-field gradient (PFG) diffusion NMR [2,20 21,22,23,24,25,26]. However, the general fractional diffusion has only been investigated in PFG NMR studies [23,24,25] and not in the relaxation NMR study. In NMR relaxation, researchers have only used the first two kinds of anomalous diffusions: the time-fractional diffusion and the space-fractional diffusion [9,26,27]. As the time-fractional diffusion and space-fractional diffusion can be viewed as two specific cases of the general fractional diffusion, developing a new theoretical treatment based on the general fractional diffusion equation could provide a broader view for NMR relaxation and help researchers to better understand dynamic processes in macromalecular systems.

This paper attempts to use the general fractional diffusion [16,17] to give correlation time functions for analyzing NMR relaxation experiments. Both the molecular rotational diffusion and molecular translational diffusion are modeled by the general fractional diffusion with $\{0 < \alpha, \beta \leq 2\}$. The general fractional diffusion equation yields a Mittag-Leffler type correlation function based on the fractional derivative. At the small-time values, the Mittag-Leffler function can be approximated as the stretched exponential function, which agrees with the commonly used KWW function [6,7]. The correlation time functions can be used to calculate NMR relaxation time that is affected by various interactions such as dipolar coupling, quadrupolar coupling, nuclear Overhauser enhancement (NOE), etc. To verify the obtained theoretical results, the obtained NMR relaxation time equation for quadrupolar coupling is used to fit reported experimental results from the literature [5]. The current fitting method is more convenient than that of the modified-KWW equation method [5], and there is good agreement between the theoretical prediction and experimental results. Additionally, it is found that the segmental dynamics under anomalous diffusion could obey a different Vogel-Tamman-Fulcher (VTF) temperature dependence than that traditionally used by the modified KWW function. The results obtained here also agree with various reported results [8,9,11,27].

**2. Theory**

**2.1 Correlation time and spectral density based on time-space fractional diffusion**

NMR relaxation is about the recovery the population distribution in a spin system after being alerted by a perturbed field. The transition speed determines the rate of the NMR relaxation process in a spin system, which can be calculated by the correlation time function of molecular motion such as rotational or translational motions [1]. Non-mono-exponential correlation functions such as KWW correlation function [6,7] or Mittag-Leffler function [8-12] are found in complicated systems, which could be described by the time-space fractional diffusion equation based on the fractional derivative [16,17] as

$$_{t}D_{*}^{\alpha}P = D_{f}\Delta^{\beta/2}P, \qquad (1)$$

where $0 < \alpha, \beta \leq 2$, $D_f$ is the fractional diffusion coefficient with units of m$^\beta$/s$^\alpha$, $_{t}D_{*}^{\alpha}$ is the Caputo fractional

derivative defined by

$$_tD_*^\alpha f(t) := \begin{cases} \frac{1}{\Gamma(m-\alpha)} \int_0^t \frac{f^{(m)}(\tau)d\tau}{(t-\tau)^{\alpha+1-m}}, m-1 < \alpha < m, \\ \frac{d^m}{dt^m}f(t), \alpha = m, \end{cases} \quad (2)$$

and $\Delta^{\beta/2}$ is the symmetric Riesz space-fractional derivative operator [23,28,29]. When $\alpha = 1, \beta = 2$, Eq. (1) reduces to the normal diffusion.

### 2.1.1 Rotational diffusion

The most common model for the normal molecule rotational motion is the normal isotropic rotational diffusion along a spherical surface proposed by Debye. Unlike the normal rotational diffusion, anomalous rotational diffusion along the spherical surface with $0 < \alpha, \beta \leq 2$ will be considered here. From the fractional diffusion Eq. (1), the anomalous rotational diffusion equation could be written as

$$_tD_*^\alpha P = \frac{D_{fr}}{a^\beta} \Delta_s^{\beta/2} P, \quad (3)$$

where $\Delta_s^{\beta/2}$ is the space-fractional derivative operator in spherical coordinate, $a$ is the radius, and $D_{fr}$ is the rotational diffusion coefficient. By separating variables, the propagator in Eq. (3) can be written as

$$P(\Omega, t) = \sum T_l^m(t) Y_l^m(\Omega). \quad (4)$$

Substituting $P(\Omega, t)$ into Eq. (2) gives a pair of eigenequations [28,29]

$$\begin{cases} _tD_*^\alpha T_l^m(t) = -D_{fr}[l(l+1)]^{\beta/2} T_l^m(t), \\ \Delta_s^{\beta/2} Y_l^m(\Omega) = -[l(l+1)]^{\beta/2} Y_l^m(\Omega). \end{cases} \quad (5)$$

From Eq. (5), we have [30]

$$T_l^m(t) = T_l^m(0) E_{\alpha,1}\left(-\frac{t^\alpha}{t_l}\right), \quad (6)$$

where

$$t_l = \frac{1}{\frac{D_{fr}}{a^\beta}[l(l+1)]^{\beta/2}}. \quad (7)$$

If the initial condition $P(\Omega, 0) = \frac{1}{4\pi}\delta(\Omega - \Omega_0)$ is used, $T_l^m(0) = \frac{1}{4\pi} Y_l^m(\Omega_0)$, because $\delta(\Omega - \Omega_0) = \sum Y_l^{m*}(\Omega_0) Y_l^m(\Omega)$. Therefore, the probability distribution function for anomalous rotational diffusion is

$$P(\Omega, t) = \frac{1}{4\pi} \sum_{l,m} Y_l^{m*}(\Omega_0) Y_l^m(\Omega) E_{\alpha,1}\left(-\frac{t^\alpha}{t_l}\right), \quad (8)$$

The correlation function can be calculated by [1]

$$G(t) = \frac{1}{4\pi} \iint F^*(\Omega) F(\Omega_0) \sum_{l,m} Y_l^{m*}(\Omega_0) Y_l^m(\Omega) E_{\alpha,1}\left(-\frac{t^\alpha}{t_l}\right) d\Omega d\Omega_0, \quad (9)$$

where $F(\Omega)$ is space-dependent lattice operator of NMR relaxation [1].

For dipole-dipole interaction, the lattice operators $F^{(0)}(\Omega)$, $F^{(1)}(\Omega)$, and $F^{(2)}(\Omega)$ can be expressed based on normalized spherical harmonic as [1]

$$F^{(0)}(\Omega) = -\frac{1}{r^3}\sqrt{\frac{16\pi}{5}} Y_2^{(0)}(\Omega)$$

$$F^{(1)}(\Omega) = \frac{1}{r^3}\sqrt{\frac{8\pi}{15}} Y_2^{(1)}(\Omega)$$

$$F^{(2)}(\Omega) = \frac{1}{r^3}\sqrt{\frac{32\pi}{15}} Y_2^{(2)}(\Omega), \quad (10)$$

where $r$ is the distance between the two interacting spins. Substituting $F^{(i)}(\Omega), i = 0,1,2$ into Eq. (9), we get

$$G^{(0)}(t) = \frac{1}{r^6}\frac{4}{5} E_{\alpha,1}(-t^\alpha/t_2)$$

$$G^{(1)}(t) = \frac{1}{r^6}\frac{2}{15}E_{\alpha,1}(-t^\alpha/t_2)$$
$$G^{(2)}(t) = \frac{1}{r^6}\frac{8}{15}E_{\alpha,1}(-t^\alpha/t_2), \tag{11}$$

where
$$t_2 = \frac{1}{6^{\frac{\beta}{2}} \times \frac{D_{f_r}}{\alpha^\beta}}. \tag{12}$$

The Fourier transform of the correlation function gives the spectral density,
$$J^{(i)}(\omega) = 2\int_0^\infty G^{(i)}(t)e^{-i\omega t}dt, \; i = 0,1,2. \tag{13}$$

If we set $I_r(\omega) = \int_0^\infty E_{\alpha,1}(-t^\alpha/t_2)e^{-i\omega t}dt$, based on the Fourier transform of MLF [16,17], we have
$$I_r(\omega) = \int_0^\infty E_{\alpha,1}(-t^\alpha/t_2)e^{-i\omega t}dt = \frac{\omega^{\alpha-1}\tau_r^\alpha \sin(\pi\alpha/2)}{1+2(\omega\tau_r)^\alpha \cos(\pi\alpha/2)+(\omega\tau_r)^{2\alpha}}, \tag{14}$$

where
$$\tau_r = \left(\frac{1}{6^{\frac{\beta}{2}} \times \frac{D_{f_r}}{\alpha^\beta}}\right)^{1/\alpha}, \tag{15}$$

and
$$J^{(0)}(\omega) = \frac{1}{r^6}\frac{4}{5} \times 2I_r(\omega), \; J^{(1)}(\omega) = \frac{1}{r^6}\frac{2}{15} \times 2I_r(\omega), \; J^{(2)}(\omega) = \frac{1}{r^6}\frac{8}{15} \times 2I_r(\omega), \tag{16}$$
which can be used to calculate the NMR relaxation time in section 2.2.

### 2.1.2 Translational diffusion

In some spin systems, the molecules' translational motion needs to be considered for NMR relaxation such as the relaxation resulting from dissolved paramagnetic impurities [1,9,27]. For anomalous translational diffusion, the Fourier transform of the fractional diffusion equation, Eq. (1), gives
$$_tD_*^\alpha P(k,t) = -2D_{f_t}k^\beta P(k,t), \tag{17}$$

where $2D_{f_t}$ is assumed to be the relative fractional translational diffusion coefficient of the two interactive spins which have the same diffusion coefficient, $D_{f_t}$, and $P(k,t) = \int_{-\infty}^\infty P(r,t)e^{ikr}dr$. Solving Eq. (17) yields
$$P(k,t) = E_{\alpha,1}\left(-2D_{f_t}k^\beta t^\alpha\right). \tag{18}$$

From Fourier transform, we have
$$P(r,t) = \frac{1}{2\pi}\int_{-\infty}^\infty P(k,t)e^{ikr}dk$$
$$= \frac{1}{2\pi}\int_{-\infty}^\infty E_{\alpha,1}\left(-2D_{f_t}k^\beta t^\alpha\right)e^{ikr}dk. \tag{19}$$

By substituting $P(r,t)$ into Eq. (7), the correlation time
$$G(t) = \frac{1}{8\pi^2}\iiint F^*(\Omega)F(\Omega_0)\, E_{\alpha,1}\left(-2D_{f_t}k^\beta t^\alpha\right)e^{ikr}dk d\Omega d\Omega_0. \tag{20}$$

For dipole-dipole interaction can be calculated following the strategy on Ref. [1] as
$$G^{(i)}(t) = \varepsilon^{(i)}\frac{N}{d^3}\int_0^\infty \left[J_{\frac{3}{2}}(u)\right]^2 E_{\alpha,1}\left(-\frac{2D_{f_t}u^\beta t^\alpha}{d^\beta}\right)\frac{du}{u}, \tag{21}$$

where
$$\varepsilon^{(0)} = \frac{48\pi}{15},\; \varepsilon^{(1)} = \frac{8\pi}{15},\; \varepsilon^{(2)} = \frac{32\pi}{15},$$

$d$ is the distance, $u = kd$, $N$ is the density of spins, and $J_{\frac{3}{2}}$ are Bessel function [1]. When $\alpha = 1, 0 < \beta \leq 2$, $E_{\alpha,1}\left(-\frac{2D_{f_t}u^\beta t^\alpha}{d^\beta}\right) = \exp\left(-\frac{2D_{f_t}u^\beta t}{d^\beta}\right)$, and the corresponded correlation function for dipole-dipole interaction is

$$G^{(i)}(t) = \varepsilon^{(i)} \frac{N}{d^3} \int_0^\infty \left[J_{\frac{3}{2}}(u)\right]^2 \exp\left(-\frac{2D_{f_t} u^\beta t}{d^\beta}\right) \frac{du}{u}, \alpha = 1, 0 < \beta \leq 2 , \tag{22}$$

which agrees with the correlation function obtained based on Levy flight reported in reference [27]. The Fourier transform of the correlation time function [16,17], Eq. (17), gives the spectral function

$$\begin{aligned} J^{(i)}(\omega) &= 2 \int_0^\infty G^{(i)}(t) e^{i\omega t} dt \\ &= \frac{\varepsilon^{(i)} N \tau_t^{1-\alpha}}{D_{f_t} d^{3-\beta} (\omega \tau_t)^{1-\alpha}} \int_0^\infty \frac{\sin(\pi\alpha/2) u^{\beta-1}}{u^{2\beta} + 2(\omega\tau_t)^\alpha \cos(\pi\alpha/2) + (\omega\tau_t)^{2\alpha}} \left[J_{\frac{3}{2}}(u)\right]^2 du, \end{aligned} \tag{23}$$

where

$$\tau_t = \left(\frac{d^\beta}{2D_{f_t}}\right)^{1/\alpha}. \tag{24}$$

When $\beta = 2$, the spectral density in Eq. (19) is consistent with that reported in Ref. [9]. The spectral density functions obtained here will be used to calculate the NMR relaxation time in the next section, section 2.2.

## 2.2 Calculate NMR relaxation

The NMR relaxation time can be determined by the correlation functions $G(t)$, or equivalently the related spectral density functions $J(\omega)$.

The spin-lattice relaxation time $T_1$ and spin-spin relaxation time $T_2$ of like spins affected by dipolar coupling can be calculated by the following equations [1]:

$$\frac{1}{T_1} = \frac{3}{2} \gamma^4 \hbar^2 I(I+1) [J^{(1)}(\omega) + J^{(2)}(2\omega)], \tag{25}$$

$$\frac{1}{T_2} = \gamma^4 \hbar^2 I(I+1) \left[\frac{3}{8} J^{(2)}(2\omega) + \frac{15}{4} J^{(1)}(\omega) + \frac{3}{8} J^{(0)}(0)\right]. \tag{26}$$

Eqs. (25) and (25) can be used to calculate relaxation time resulted from either the rotational diffusion or the translational diffusion.

In macromolecular systems, the chain segmental rotation motion is the dominant factor affecting the NMR relaxation. Under the influence of rotational motion, NMR relaxation time affected by other interactions such as dipolar coupling between unlike spins and quadrupolar interaction, etc. can also be obtained according to Ref. [1]. For $^{13}C$ nucleus, C-H dipole-dipole interaction gives a $^{13}C$ spin-lattice relaxation time

$$\frac{1}{T_1} = \frac{\gamma_C^2 \gamma_H^2 \hbar^2}{10 r^6} n [I_r(\omega_H - \omega_c) + 3 I_r(\omega_c) + 6 I_r(\omega_H - \omega_c)] , \tag{27}$$

where $n$ is the number of proton nuclei directly bonded to the carbon nucleus, and $I_r(\omega)$ is defined by Eq. (14).

For deuterium, the quadrupolar relaxation [1] expression is

$$\frac{1}{T_1} = \frac{3}{80} \left(1 + \frac{\eta^2}{3}\right) \left(\frac{e^2 qQ}{\hbar}\right)^2 [I_r(\omega) + 4 I_r(2\omega)], \tag{28}$$

where is $\eta$ symmetric parameter of the quadrupolar moment, and $I_r(\omega)$ is defined by Eq. (14).

## 3. Results and discussion

The general correlation function and related spectral density functions are derived based on the time-space fractional diffusion equation, $_t D_*^\alpha P = D_f \Delta^{\beta/2} P$, where $0 < \alpha, \beta \leq 2$. The obtained time correlation functions are Mittag-Leffler type functions $G^{(i)}(t)$ described by Eqs. (9), (11) and (20). For relaxation due to dipolar coupling, the fractional rotational diffusion has $G^{(i)}(t) \propto E_{\alpha,1}(-D_{f_r}[l(l+1)]^{\beta/2} t^\alpha)$, while the fractional translational diffusion has $G^{(i)}(t) = \varepsilon^{(i)} \frac{N}{d^3} \int_0^\infty \left[J_{\frac{3}{2}}(u)\right]^2 E_{\alpha,1}\left(-\frac{2D_{f_t} u^\beta t^\alpha}{d^\beta}\right) \frac{du}{u}$; their corresponding spectral densities are described by Eqs. (16) and (23) for rotational diffusion and translational diffusion, respectively. When $\alpha = 1, \beta = 2$, fractional diffusion reduces to normal diffusion, and all the correlation time and spectral densities reported here are reduced to the reported normal diffusion results in Ref. [1].

These results are consistent with various reported results. For the time-fractional diffusion, $0 < \alpha, \leq 2, \beta = 2$, the rotational correlation time $E_{\alpha,1}(-D_{fr}l(l+1)t^\alpha)$ is consistent with the correlation time reported in [26]; the translational correlation time $\varepsilon^{(i)}\frac{N}{d^3}\int_0^\infty \left[J_{\frac{3}{2}}(u)\right]^2 E_{\alpha,1}\left(-\frac{2D_{ft}u^2t^\alpha}{d^2}\right)\frac{du}{u}$ will give the spectral density $\frac{\varepsilon^{(i)}N\tau_t^{1-\alpha}}{D_{ft}d(\omega\tau_t)^{1-\alpha}}\int_0^\infty \frac{\sin(\pi\alpha/2)u}{u^4+2(\omega\tau_t)^\alpha\cos(\pi\alpha/2)+(\omega\tau_t)^{2\alpha}}\left[J_{\frac{3}{2}}(u)\right]^2 du$ that agrees with the spectral density reported in Ref. [9]. On the other hand, for the space-fractional diffusion with $\alpha = 1, 0 < \beta \leq 2$, $E_{\alpha,1}\left(-\frac{2D_{ft}u^\beta t^\alpha}{d^\beta}\right) = \exp\left(-\frac{2D_{ft}u^\beta t}{d^\beta}\right)$, the translational correlation time agrees with the results reported in Ref. [27].

In the time-space fractional diffusion equation, Eq. (1), the Caputo fractional derivative ${}_tD_*^\alpha$ is used, providing a Mittag-Leffler type correlation function. In literature, there are other types of time fractional derivative such as time fractal derivative [31,32,33], $\frac{\partial}{\partial t^\alpha}$. The fractal derivative is built based on the Hausdorff derivative, which is defined as $\frac{\partial P^{\beta/2}}{\partial t^\alpha} = \lim_{t_1 \to t}\frac{P^{\beta/2}(t_1)-P^{\beta/2}(t)}{t_1^\alpha-t^\alpha}$ [31,32]. If the time fractal derivative $\frac{\partial}{\partial t^\alpha}$ instead of ${}_tD_*^\alpha$ is used in building a time-space fractional diffusion equation $-\frac{\partial}{\partial t^\alpha}P = D_f\Delta^{\beta/2}P$, different correlation functions will be given. The MLFs, $E_{\alpha,1}\left(-\frac{t^\alpha}{\tau_l}\right)$ in Eq. (8) and $E_{\alpha,1}\left(-\frac{2D_{ft}u^\beta t^\alpha}{d^\beta}\right)$ in Eq. (16) will be replaced by SEFs, $\exp\left(-\frac{t^\alpha}{\tau_l}\right)$ and $\exp\left(-\frac{2D_{ft}u^\beta t^\alpha}{d^\beta}\right)$, respectively. This coincides with the familiar KWW function. When $\alpha = 1, \beta = 2$, both the MLFs and SEFs reduce to a mono-exponential function. At small x values, $E_{\alpha,1}(-x) \approx \exp[-x/\Gamma(1+\alpha)]$, but at large $x$ values, the MLF decays slower than SEF.

The NMR relaxation experiments for macromolecular systems are often variable temperature (VT) experiments where a set of relaxation time values are observed in a range of temperatures. The VT experiments provide not only a series of relaxation time values necessary for data fitting, but also a temperature dependent dynamic behavior that is crucial for material application in different temperature conditions. For KWW function $\exp\left(-\left(\frac{t}{\tau_r}\right)^\alpha\right)$, it is often assumed that the segmental dynamics obeys a Vogel-Tamman-Fulcher (VTF) temperature dependence: $\tau_r = \tau_\infty \times 10^{\frac{B}{T-T_0}}$. However, in Eq. (15), $\tau_r = \left(\frac{1}{6^{\frac{\beta}{2}} \times \frac{D_{fr}}{\alpha\beta}}\right)^{1/\alpha}$, which implies $(\tau_r)^\alpha \propto D_{fr} \propto \langle\tau^\alpha\rangle$ because $D_{fr} \propto \frac{\langle|z|^\beta\rangle}{\langle\tau^\alpha\rangle}$ [16,17]. If $\langle\tau^\alpha\rangle$ is assumed to be proportional to $\tau_\infty \times 10^{\frac{B}{T-T_0}}$, we may have a different VTF dependence for anomalous rotational diffusion:

$$(\tau_r)^\alpha = \tau_\infty \times 10^{\frac{B}{T-T_0}}; \qquad (29)$$

similarly, for anomalous translational diffusion:

$$(\tau_t)^\alpha = \tau_\infty \times 10^{\frac{B}{T-T_0}}. \qquad (30)$$

Eqs. (29) and (30) are different from the traditional VTF dependences:

$$\tau_r = \tau_\infty \times 10^{\frac{B}{T-T_0}}, \qquad (31)$$

and

$$\tau_t = \tau_\infty \times 10^{\frac{B}{T-T_0}}. \qquad (32)$$

Currently, it is still not clear which set of equations are right. Both equations will be used to fit the experimental data in the next paragraph.

The theoretical results can be applied to fit the experimental data. The Mittag-Leffler type correlation function could replace the KWW function in fitting NMR experimental data. The form of the time correlation function reported in the literature can become complicating [4,5]. For instance, a modified KWW (mKWW) function $G(t) = a_{lib}\exp\left[-\frac{t}{\tau_{lib}}\right] + (1-a_{lib})\exp\left[-\left(\frac{t}{\tau_r}\right)^\alpha\right]$ is used in the analysis of the spin-lattice relaxation in polymer systems in Ref. [5], which is a combination of a mono-exponential function and a KWW function. Instead, a single Mittag-Leffler type function could be used to fit the NMR experimental data. In Figure 1, Eq. (28) is used to fit the experimental data taken from Ref. [5], which investigates [2]H spin-lattice relaxation of deuterium-labeled head-to-head poly(propylene) (hhPP) in a blend consisting of 70 % polyisobutylene (PIB) and 30 % hhPP [5]. The fitting parameters are listed in Table 1. The fitting given by the Mittag-Leffler type function is comparably similar to the fitting provided by the mKWW function reported in Ref. [5]. However, the fitting by Mittag-Leffler type function only uses four parameters, $\alpha$, $\tau_\infty$, $B$, and $T_0$, while the mKWW function needs six parameters, $\alpha$, $\tau_\infty$, $B$, $T_0$, $a_{lib}$, and $\tau_{lib}$. Two different VTF temperature dependencies: Eq. (29) and Eq. (31) are used to yield two set of fitting curves. Both the VTF temperature dependencies can fit the reported experimental data taken from Ref. [5]. Because the difference between the two predicted curves is not obvious, it is difficult to conclude whether Eq. (29) or Eq. (31) is the right Vogel-Tamman-Fulcher equation. This requires further research efforts. Both fittings obtain the same $\alpha$ value equaling 0.6, which is slightly higher than $\alpha = 0.5$ reported in Ref. [5]. Additionally, a modified Mittag-Leffler type function $a_{lib} \exp\left[-\frac{t}{\tau_{lib}}\right] + (1 - a_{lib}) E_{\alpha,1}\left(-\frac{t^\alpha}{t_l}\right)$ is used to fit the reported experimental data [5], but the most appropriate fitting always has $a_{lib} = 0$, namely, no liberational motion is needed in the fitting. This may be reasonable because the Mittag-Leffler type correlation function could include both the slow motion and fast liberational motion.

**Table 1:** Fitting parameters

| Dynamic Mode | A | $\tau_{inf}$ (ps) | B (K) | $T_0$ (K) |
|---|---|---|---|---|
| Eqs. (28) and (29) | 0.60 | 940 | 975 | 110 |
| Eqs. (28) and (31) | 0.60 | 0.03657 | 910 | 172 |
| mKWW parameters taken from Ref. [5] $\tau_{lib} = 0.1$ ps, $a_{lib} = 0.07$ | 0.5 | 0.1 | 730 | 188 |

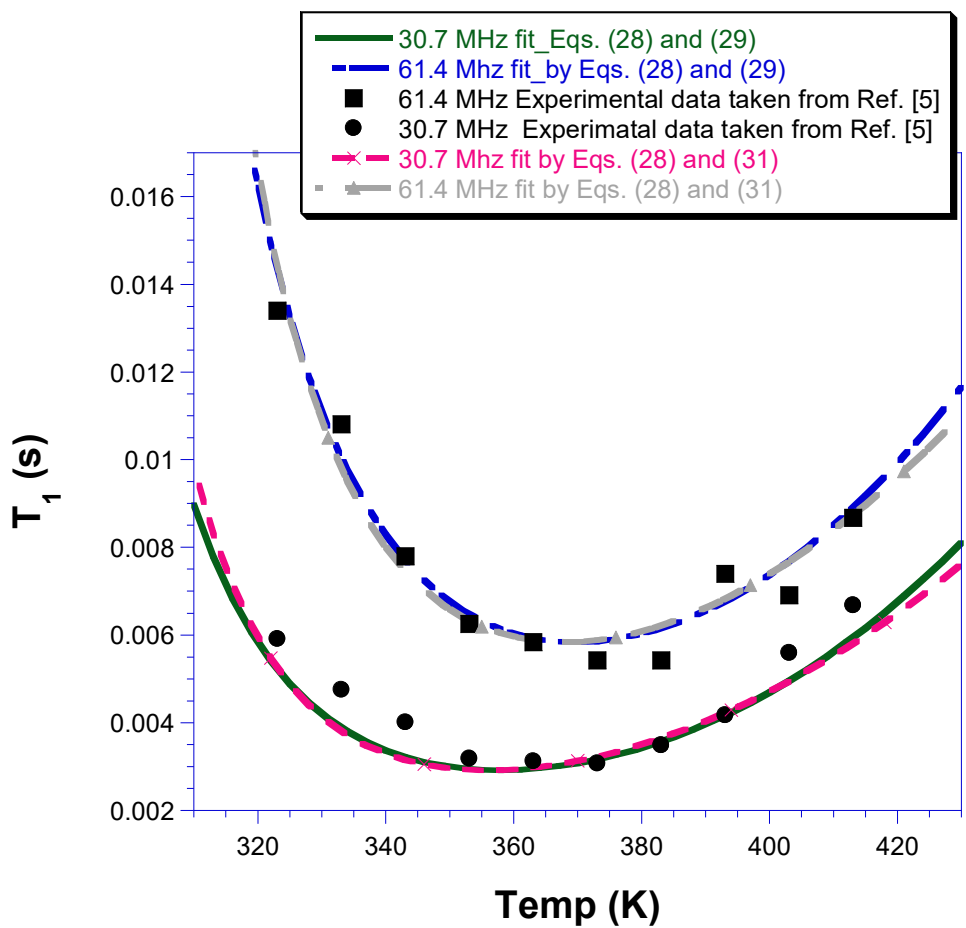

**Figure** 1. Fit the deuterium NMR spin-lattice relaxation by the quadrupolar coupling relaxation equation, (28). The experimental data are taken from Ref. [5]. Two different VTF temperature dependence equations, (29) and (31) are also compared in the two set of fitting curves.

The translational motion and rotational motion are related to each other. Generally, a molecule with fast translational motion rotates quickly. For normal diffusion, Refs. [34, 35] suggested that $D_t \tau_r = 2r_s^2/9$ based on the combination the Stokes-Einstein and Debye-Stokes-Einstein equations. A similar expression $D_{f_t} \tau_r^\alpha \propto 2|r_s|^\beta/[9\Gamma(1+\alpha)]$ may be assumed as a possible substitution to relate the rotational motion and translational motion in an anomalous diffusion.

The focus of this paper is NMR relaxation, and the results here can be combined with PFG diffusion NMR and other methods to provide a broader view of the molecular dynamic processes in a complicated system.